\documentclass[aps,prc,showpacs,nofootinbib,twocolumn]{revtex4}

\usepackage{epsfig}
\usepackage{graphicx,amsmath}
\newcommand\ba{\begin{eqnarray}}
\newcommand\ea{\end{eqnarray}}

\newcommand{\be}{\begin{equation}}
\newcommand{\ee}{\end{equation}}

\newcommand{\bas}{\begin{eqnarray*}}
\newcommand{\eas}{\end{eqnarray*}}
\begin{document}
\title{\bf \large Search for Low Mass Exotic mesonic structures. Part II: attempts to understand the experimental results}

\author{
B. Tatischeff$^{1,2}$\thanks{e-mail : tati@ipno.in2p3.fr}\\
$^{1}$CNRS/IN2P3, Institut de Physique Nucl\'eaire, UMR 8608, Orsay, F-91405\\
$^{2}$Univ. Paris-Sud, Orsay, F-91405, France\vspace{3.mm}\\
E. Tomasi-Gustafsson$^{3}$\thanks{e-mail : etomasi@cea.fr}\vspace*{1.mm}\\
$^{3}$DAPNIA/SPhN, CEA/Saclay\\ 91191 Gif-sur-Yvette Cedex, France}

\pacs{13.60.Le, 14.40.Cs, 14.80.-j}

\vspace*{1cm}
\begin{abstract}
 Our previous paper, part  I of the same study, shows the different experimental spectra used to conclude on the genuine existence of 
narrow, weakly excited mesonic structures, having masses below and a little above the pion (M=139.56~MeV) mass. This work \cite{previous} was instigated by the 
observation, in the  $\Sigma^{+}$ disintegration:
$\Sigma^{+}\to$pP$^{0}$, P$^{0}\to\mu^{-}\mu^{+}$ \cite{park}, of a narrow range of dimuon masses. The authors conclude on the existence of a neutral intermediate state P$_{0}$, with a mass M=214.3~MeV $\pm$ 0.5~MeV. 
We present here some attempts to understand the possible nature of the structures
observed in part I.
\end{abstract}
\maketitle
\section{Introduction} 
In part I of the same study \cite{previous}, several spectra were presented, showing narrow and weakly excited structures, having masses below and just above the pion mass (M=139.56~MeV). These data are mainly missing mass precise spectra of the  pp$\to$ppX reaction studied at SPES3 (Saturne). Different selected results from COSY, Celsius, MAMI, and JLAB Hall A, Hall B,  and Hall C were also shown, which confirm the genuine existence of the structures. The statistical confidence was often large. In some cases, this confidence is not large, but several structures at the $\approx$ same masses were observed.

These masses are M=62~MeV, 80~MeV, 100~MeV, 181~MeV, 198~MeV, 215~MeV, 227.5~MeV, and 235~MeV, although the last one may be uncertain, since determined by only three data, and being located at the limit of the spectra. They are shown in Fig.~1. A few points, located around M=75~MeV, may be thought as being not resolved structures. Indeed when they are extracted none of the structures at M=100 or M=62~MeV is observed. However the symmetry of the masses reported in Fig.~1, may be considered as an indication of their genuine existence. 
  \begin{figure}[!h]
\begin{center}                                                          
\scalebox{.48}[.48]{
\includegraphics[bb=20 20 530 530,clip=]{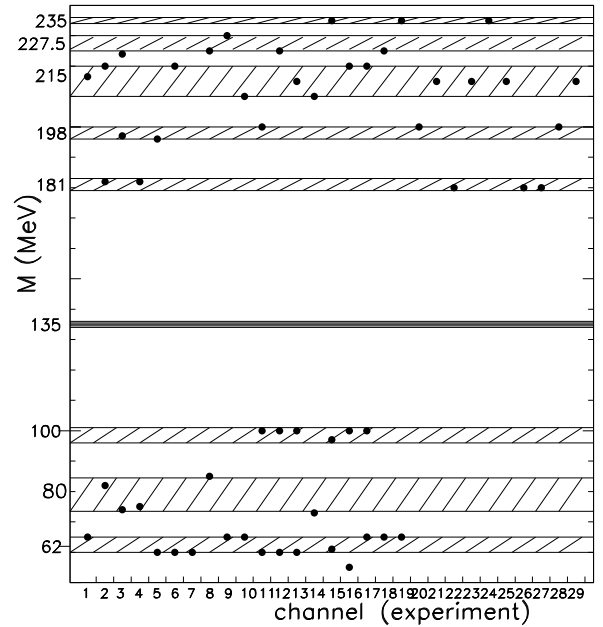}}
\caption{Masses of the weakly excited structures extracted from several experiments.}
\label{fig11}
\end{center}
\end{figure}
\section{Discussion}
\subsection{Possible mass relation between mesons and baryons }
Narrow structures in dibaryons \cite{bt}, in baryons \cite{bt2}, and in mesons at larger masses \cite{jy} \cite{bt1} were already previously observed . Mesonic structures appeared in the missing mass of the same reaction, pp$\to$ppX but in different kinematical conditions from those discussed here.
The sequence of the presently discussed low mass exotic mesons, reproduce fairly well the sequence of the low mass exotic baryons \cite{bt2}. 
Indeed the three first narrow baryonic structure masses are M=1004~MeV, 1044~MeV, and 1094~MeV. We note that the first baryonic mass differences are: M=1004-939=65 MeV,  1044-939=105~MeV, and 1094-939=155~MeV. The two first values agree with the masses
extracted in the present study (and the third one is to compare to the meson mass). Such property may indicate a relation between baryons and mesons. Already long time ago, the possibility to calculate the baryon spectrum, starting from the meson one was considered \cite{richard}. The shifts between adjacent masses, close to $\Delta$M$\approx$20~MeV or 40~MeV, compare favourably to the model \cite{walcher} which suggests the existence of a "genuine" Goldstone Boson with a mass close to 20~MeV.

We observe mass differences close to $\Delta$M$\approx$35~MeV. The mass difference between the three lowest masses extracted before, is close to 
$\Delta$M=17.5~MeV. A mass gap  of 
\vspace*{-4.mm}
$$\Delta M=35~MeV=\displaystyle\frac{1}{2} \displaystyle\frac{m_e}{\alpha},$$
where $m_e$ is the electron mass and $\alpha$ is the fine structure constant,  was observed between several narrow hadronic exotic masses \cite{bt}. Such gaps for leptons, mesons, and baryons were discussed since long time \cite{macgregor} and recently discussed again \cite{palazzi}.The mass difference observed here equals half this value. We notice that the first excited state of the light and stable nucleus $^4$He, is close to $\Delta$M$\approx$20~MeV, not far from 17.5~MeV. The shift between both values could eventually be associated with residual interactions between several nucleons. 
\subsection{Regge-like trajectories}
In order to support the hadronic nature of these low mass mesonic structures, we study the Regge-like trajectories of all narrow exotic mesonic structure masses extracted \cite{jy} \cite{bt1} and those of the present work 
\vspace*{0.mm}
\begin{equation}
n=a+b M^{2}. 
\label{eq:eqm}
\end{equation}
This is shown in Fig.~2 where filled circles correspond to experimental structure masses and empty squares to not observed  masses. These masses follow three straight lines, the two inflexion points corresponding to one and two pion masses. We name them Regge-like trajectories, since the ordinate "n" is arbitrary (as "a") and corresponds to an unknown quantum number and not to the spin. The slopes "b" of the straight lines are much larger than the slope of "classical" 
(PDG) mesons slope \cite{pdg} which is close to 0.9~GeV$^{-2}$ when determined in the range $\rho$(770)$\le$m$\le$f$_{6}$(2510) \cite{pdg}. Namely their values, for the three straight lines from small to larger mesonic masses, are: 390, 149.7, and 32.5~GeV$^{-2}$. We notice that these four slopes, including the PDG meson slope vary continuously. This is shown in Fig.~3 where the slope is reported versus the mean mesonic mass range (MM).
\begin{figure}[!h]
\begin{center}                                                          
\scalebox{.48}[.48]{
\includegraphics[bb=20 20 530 530,clip=]{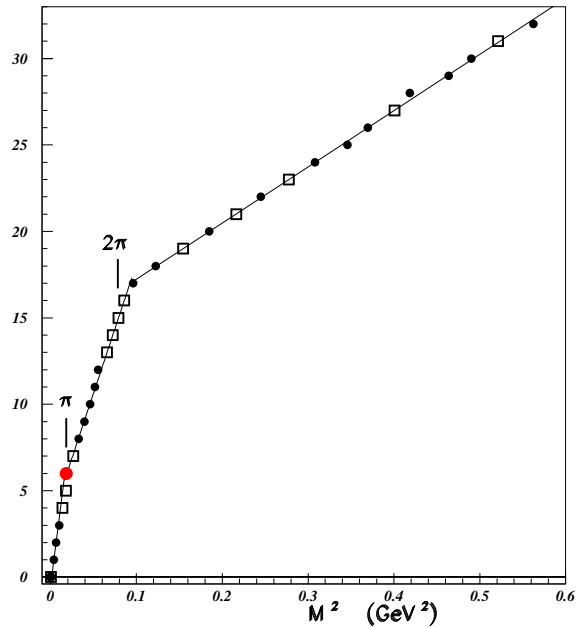}}
\caption{Regge-like trajectory of all narrow structure mesonic masses observed in this work and in \protect\cite{jy,bt1}. Solid circles (empty squares) represent observed(expected) resonances.}
\label{fig2}
\end{center}
\end{figure}
\begin{figure}[!h]
\begin{center}                                                          
\scalebox{.48}[.58]{
\includegraphics[bb=35 265 523 573,clip=]{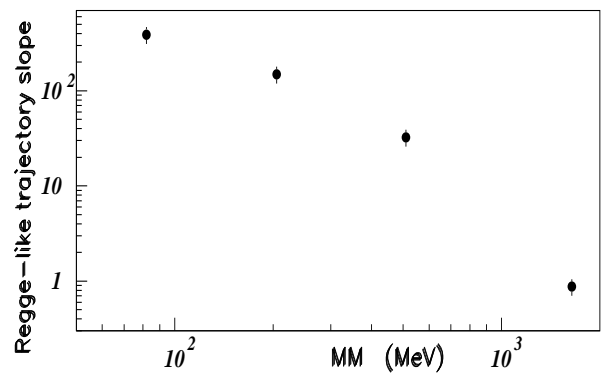}}
\caption{Slopes of the Regge-like trajectories versus the mean mesonic mass of each straight line range.}
\label{fig3}
\end{center}
\end{figure} 
\subsection{Kaluza-Klein mass formula}
It has been shown that the Kaluza-Klein formula \cite{arkhi} predicts rather well the masses of the narrow exotic baryons experimentally observed  \cite{bota} and also predicts although a little less well, the masses of the narrow dibaryons experimentally observed. We use the same relation for one particle:
\vspace*{-0.mm}
$$ m_{n}^{2}=m_{0}^{2} + n^{2}/R^{2} $$

\hspace{-3.mm}where R$^{-1}$=41.481~MeV is the fundamental scale parameter and "n" the sequence of integer numbers. This second term describes the contribution to the mass of the extra dimensions.
If we choose - arbitrarily - m$_{0}$=63.3~MeV, adjusted to obtain m=m$_{\pi}^{ \hspace{1.mm}+}$ for n=3,
we get the calculated masses shown in Fig.~4. These values compare quite well with several experimental masses of narrow low mass mesonic structures. For instance,  up tp M=190~MeV, the number of calculated states and their masses are well reproduced.
\begin{figure}[!h]
\begin{center}                                                          
\scalebox{1}[0.8]{
\includegraphics[bb=60 280 315 516,clip=]{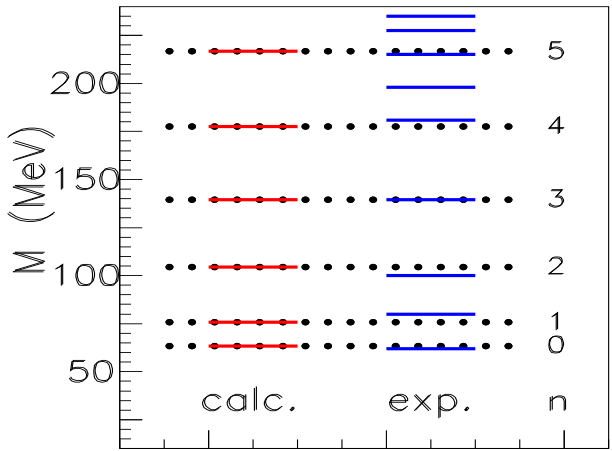}}
\caption{Narrow meson low masses calculated using the Kaluza Klein formula (n=0,1,2,3,4,5).}
\label{fig4}
\end{center}
\end{figure}
\subsection{What kind of particles ?}
The possibility to explain the 511~KeV $\gamma$ ray line observed by the INTEGRAL satellite from the "galactic bulge", was discussed in relation with spin 0 or 1 boson particles, candidates to dark matter \cite{fayet}. Indeed as said by the author of 
ref.\cite{hooper}, "It has long been understood that weakly interacting particles with masses smaller than a few GeV (but larger than $\approx$1~MeV) are expected to be overproduced in the early universe relative to the dark matter abundance". The author of ref.\cite{fayet} applied his calculations to dark matter particle of mass either M=1~MeV, 10~MeV, 100~MeV or 1~GeV. Although the range of masses fits the masses observed in this work, we have no argument to identify them to eventual dark matter particle(s). 
\section{Conclusion}
 We have discussed the nature of  the structures discussed in the previous part I of this study \cite{previous}. Their hadronic nature is supported by the Regge-like trajectory shown in Fig.~2.
The masses of narrow exotic mesons, baryons, and dibaryons, observed before, were attributed to quark clusters \cite{bt,bt2,jy,bor}. For a review, see also
\cite{Bo74}. We suggest that the structures shown here are also due to quark clusters as the previous ones.
Since their masses are lower than the mass of two pions, they can only disintegrate through electromagnetic or weak interactions.

We thank Prof.  E.A. Kuraev for valuable discussions.

\end{document}